\documentclass[aps,prl,twocolumn,superscriptaddress,showpacs,amsmath]{revtex4}

\usepackage{graphicx,amssymb}

\begin{document}

%

\title{Mesoscopic quantum coherence in antiferromagnetic molecular clusters}

\author{O. Waldmann}
\email[Corresponding author.\\E-mail: ]{waldmann@iac.unibe.ch} \affiliation{Department of Chemistry and
Biochemistry, University of Bern, 3012 Bern, Switzerland}

\author{C. Dobe}
\affiliation{Department of Chemistry and Biochemistry, University of Bern, 3012 Bern, Switzerland}

\author{H. Mutka}
\affiliation{Institut Laue-Langevin, 6 rue Jules Horowitz, BP 156, 38042 Grenoble Cedex 9, France}

\author{A. Furrer}
\affiliation{Laboratory for Neutron Scattering, ETH Z\"urich and PSI, 5232 Villigen PSI, Switzerland}

\author{H. U. G\"udel}
\affiliation{Department of Chemistry and Biochemistry, University of Bern, 3012 Bern, Switzerland}

\date{\today}

\begin{abstract}
From inelastic neutron scattering experiments, we demonstrate quantum tunneling of the N\'eel vector in the
antiferromagnetic molecular ferric wheel CsFe8. Analysis of the linewidth of the tunneling transition
evidences coherent tunneling.
\end{abstract}


\maketitle

%

The theoretical consideration of mesoscopic quantum coherence (MQC) in small antiferromagnetic (AF) particles
a decade ago, and the prediction that it should be more robust than in ferromagnetic particles \cite{Bar90},
has stimulated intense efforts in this direction. Subsequent experimental studies on ferritin clusters
remained controversial \cite{Gid95,Tej96}. Several years ago it was suggested that AF molecular wheels, i.e.,
molecules in which an even number of magnetic metal centers are arranged in a ring, may exhibit MQC as
coherent tunneling of the N\'eel vector \cite{Chi98,Mei01}. By means of inelastic neutron scattering (INS)
experiments, we now demonstrate this MQC effect in the AF molecular wheel
[CsFe$_8L_8$]Cl$\cdot$2CHCl$_3$$\cdot$0.5CH$_2$Cl$_2$$\cdot$0.75$L$$\cdot$2.5H$_2$O with $L$ =
N(CH$_2$CH$_2$O)$_3$, or CsFe8 in short. Our finding represents the first clear demonstration of MQC in an AF
cluster, ending a decade long hunt to observe such phenomena.

\begin{figure}[b]
\includegraphics{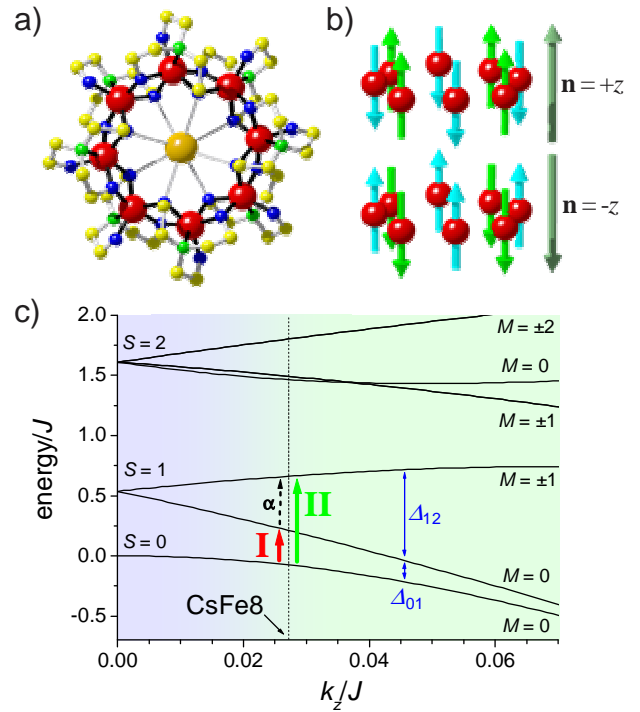}
\caption{\label{fig1} (a) Crystal structure of CsFe8 (Fe: red, O: blue, N: green, C: yellow, Cs: orange, H
atoms are omitted). (b) Classical ground-state spin configuration for the two orientations of the N\'eel
vector (dark green arrows) with $\mathbf{n} = \pm z$. (c) Low-lying energy spectrum of $\hat{H}$ in units of
$J$ for an octanuclear $s = 5/2$ wheel as function of $k_z/J$. At small anisotropies, the levels can be
classified by the spin quantum numbers $S$ and $M$. For large anisotropies, $S$ loses its significance ($M$
remains an exact quantum number), and the region of N\'eel-vector tunneling is entered (transition from blue
to green shading). The red, green, and black arrows indicate transitions I, II and $\alpha$; blue arrows the
energy gaps $\Delta_{01}$, $\Delta_{12}$.}
\end{figure}

In the CsFe8 molecule \cite{Saa97}, eight Fe(III) ions form a regular octagon with the Cs ion located in the
center of the ring [see Fig.~\ref{fig1}(a)]. The magnetic properties of AF molecular wheels, including CsFe8,
are well described by the spin Hamiltonian \cite{Pil01,XFe6,CsFe8}
\begin{equation}
 \hat{H} = J \left( \sum^{N-1}_{i=1}{ \hat{\textbf{S}}_i \cdot \hat{\textbf{S}}_{i+1} } +
\hat{\textbf{S}}_N \cdot \hat{\textbf{S}}_1 \right) - k_z \sum^N_{i=1} \hat{S}^2_{i,z}
\end{equation}
with isotropic nearest-neighbor exchange interactions, characterized by the coupling constant $J > 0$, and an
uniaxial magnetic anisotropy, of the easy-axis type, described by an on-site anisotropy with $k_z > 0$
\cite{kz,dip} ($N = 8$ is the number of spin centers, $\hat{\textbf{S}}_i$ is the spin operator of the $i$th
ion with spin $s = 5/2$ for Fe(III), and $z$ is the uniaxial anisotropy axis perpendicular to the plane of
the wheel).

%

\begin{figure}
\includegraphics{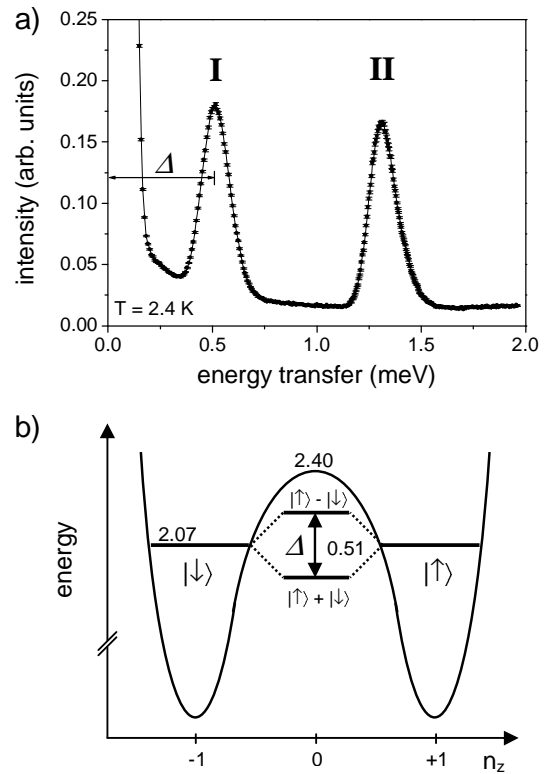}
\caption{\label{fig2} (a) Neutron energy-loss spectrum of CsFe8 at $T$ = 2.4~K ($\lambda$ = 5.0~\AA). Data
were summed over all detector banks. As discussed in the text, transition I corresponds to NVT with tunneling
gap $\Delta$. (b) Sketch of the energy diagram for NVT in CsFe8. The numbers correspond to energies in meV;
the values for the ground state energy and barrier height are relative to the bottom of the double-well
potential.}
\end{figure}

It is useful to consider first the scenarios described by $\hat{H}$ as a function of the parameter $k_z/J$.
The calculated energy spectrum is shown in Fig.~\ref{fig1}(c) \cite{FW_QT}. For the isotropic case, $k_z/J =
0$, the eigenstates may be classified by the total spin quantum numbers $S$ and $M$. In the ground state, all
the single-ion spins are aligned antiferromagnetically and the total net magnetic moment is fully
compensated, corresponding to $S = 0$. The orientation of this AF spin configuration in space is not fixed
and allows for a coherent rotation of all the spins, which can be equivalently described by the motion of the
N\'eel vector $\mathbf{n} = (\mathbf{M}_A - \mathbf{M}_B)/(2M_0)$ ($M_0 = \mu_B N s$ is the length of the
sublattice magnetization vectors $\mathbf{M}_A$ or $\mathbf{M}_B$, respectively; $\mu_B$ is the Bohr
magneton). Quantum mechanically, this rotational degree of freedom corresponds to that of a rigid rotator,
and the eigenstates above the ground state thus belong to $S = 1, 2, \ldots$ \cite{Cr8}. In the presence of
anisotropy the spin multiplets split, and for large values of $k_z/J > 0$, the $|S,M\rangle$ functions become
mixed so strongly that $S$ loses its significance. This is accompanied by a drastic change in the dynamics of
the N\'eel vector \cite{Chi98,Mei01}. The easy-axis anisotropy acts as an energy barrier favoring an
alignment of the N\'eel vector along the $z$ axis. Accordingly, for increasing values of $k_z/J$ the energy
barrier eventually overcomes the quantum mechanical zero-point motion, and the N\'eel vector becomes
localized in the $\pm z$ directions. Classically, the system is then in either of the two degenerate states
with $\mathbf{n} = \pm z$, Fig.~\ref{fig1}(b). Quantum mechanically, however, tunneling between the quantum
analogs of these two states, $|\uparrow\rangle$ and $|\downarrow\rangle$, results in the eigenstates
$|\uparrow\rangle+|\downarrow\rangle$ and $|\uparrow\rangle-|\downarrow\rangle$, which are separated in
energy by the tunneling gap $\Delta$ [see also Fig.~\ref{fig2}(b)]. In the energy spectrum,
Fig.~\ref{fig1}(c), these tunneling states correspond to the two lowest levels at large values of $k_z/J$,
where their energy separation $\Delta_{01}$ becomes smaller than the splitting of the $S = 1$ multiplet,
$\Delta_{12}$. Qualitatively, the N\'eel vector tunneling (NVT) regime is thus entered when $\Delta_{12}
\gtrsim \Delta_{01}$. In a semi-classical analysis, which is the natural and accurate starting point to
describe NVT in molecular wheels \cite{Chi98,Mei01,Hon02}, NVT is realized for $k_z/J > 4(Ns)^{-2}$ or
equivalently $S_0/\hbar > 4$, with the tunneling action given by $S_0/\hbar = NS \sqrt{2k_z/J}$ \cite{crit}.
NVT may be demonstrated experimentally by {\it simultaneously} observing the transitions from the ground
state to the first and second excited levels [transitions I and II in Fig.~\ref{fig1}(c)], giving
$\Delta_{01}$ and $\Delta_{12}$, respectively, or via Fig.~\ref{fig1}(c), $J$ and $k_z$. This allows the
important consistency check whether the condition for NVT is indeed met. In the tunneling regime,
$\Delta_{01}$ corresponds to the tunneling gap $\Delta$.

%

The CsFe8 material was prepared by the same procedure as in Ref.~\onlinecite{Saa97}, but crystallized from a
1:1 mixture of CHCl$_3$ and CH$_2$Cl$_2$ by pentane vapor diffusion. The compound crystallizes in the
monoclinic space group Pna21, and the molecules exhibit approximate C$_4$ symmetry with Fe-Fe distances of
3.142 - 3.164~\AA, Fig.~\ref{fig1}(a). The INS spectrum of CsFe8 was measured with the time-of-flight
spectrometer IN5 at the Institute Laue-Langevin (ILL), Grenoble, France. 4~g of a non-deuterated powder
sample was sealed under a helium atmosphere in a double-walled hollow Al cylinder (50~mm in height, 16~mm
external diameter, and 2~mm thickness) and inserted in a standard ILL orange $^4$He cryostat. INS spectra
were recorded at temperatures between 2.2~K and 9.7~K for incident neutron wavelengths of $\lambda$ =
5.0~{\AA} and 8.0~{\AA} (energy resolution at the elastic peak were 121~$\mu$eV and 22~$\mu$eV,
respectively). The data were corrected for detector efficiency using a vanadium standard; no further
corrections were applied.

%

Figure~\ref{fig2}(a) shows the INS spectra obtained at a temperature of $T$ = 2.4~K with $\lambda$ = 5.0~\AA.
The high quality of the data, which is outstanding for a non-deuterated molecular spin cluster, should be
noted. Two cold peaks with slight asymmetry are visible at energy transfers of 0.51(1)~meV and 1.31(1)~meV.
Their magnetic origin is unambiguously confirmed by their dependence on temperature and momentum transfer
$Q$. The two peaks clearly correspond to transitions I and II in Fig.~\ref{fig1}(c), respectively,
demonstrating that for CsFe8 $\Delta_{01}$ = 0.51(1)~meV and $\Delta_{12}$ = 0.80(1)~meV, from which $J$ =
1.78(4)~meV and $k_z$ = 0.048(1)~meV are determined. These values are in excellent agreement with those
inferred from high-field torque and magnetic susceptibility measurements \cite{CsFe8}. NVT in CsFe8 is
immediately suggested by $\Delta_{12} \gtrsim \Delta_{01}$, and unambiguously confirmed by $S_0/\hbar = 4.6 >
4$. The semi-classical analysis of NVT draws an intuitive picture \cite{Chi98,Mei01}. The height of the
energy barrier is given by $Ns^2k_z$ = 2.40~meV, and the ground-state energy by $\hbar \omega_0$ = $s \sqrt{8
J k_z}$ = 2.07~meV \cite{Mei01}. For the tunneling gap $\Delta$, no analytical result is available. However,
numerical calculations using $\hat{H}$ indicate that $16 \hbar \omega_0 \sqrt{S_0/(2\pi \hbar)}
\exp(-S_0/\hbar)$ is a good approximation \cite{size}, yielding $\Delta \approx$ 0.3~meV. In view of the
involved approximations and their exponential effect, this is in excellent agreement with the experimental
finding $\Delta$ = 0.51~meV. In summary, the energetic situation is as depicted in Fig.~\ref{fig2}(b), and
sub-barrier tunneling is clearly realized in CsFe8.

\begin{figure}
\includegraphics{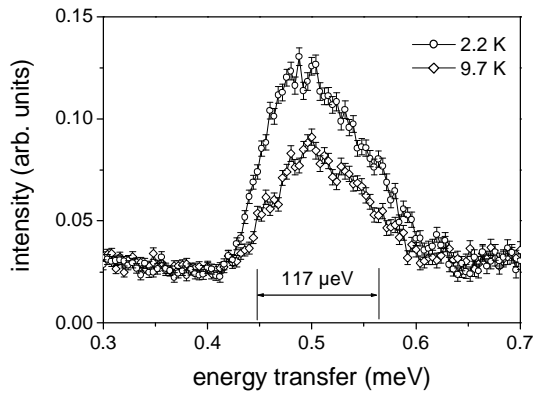}
\caption{\label{fig3} Neutron energy-loss spectrum of CsFe8 at $T$ = 2.2~K and 9.7~K ($\lambda$ = 8.0~\AA),
showing peak I with high resolution. The peak shape is independent of temperature with a FWHM as indicated.}
\end{figure}

The homogeneous linewidth of the tunneling transition, peak I, provides an estimate for the decoherence time
$\tau_\phi$ \cite{Leg87}. Peak I was studied with high resolution at $T$ = 2.2~K and 9.7~K using $\lambda$ =
8.0~\AA, see Fig.~\ref{fig3}. The full width at half-maximum (FWHM) is about 120~$\mu$eV, with some
temperature independent asymmetry in the peak shape which is evidence for considerable inhomogeneous
broadening \cite{broad}. A homogeneous width of 120~$\mu$eV would correspond to a decoherence time
$\tau_\phi$ = 11~ps. The intrinsic linewidth of transition I is obscured by the inhomogeneous broadening,
indicating a longer decoherence time, i.e., $\tau_\phi > 11$~ps. The quantity $\tau_\phi \Delta/\hbar$ is an
important figure of merit for the operation of quantum devices as it measures the number of coherent
oscillations before decoherence sets in. With the above estimate for $\tau_\phi$ and $\Delta/\hbar$ =
775~GHz, one obtains $\tau_\phi \Delta/\hbar > 8.5$ for CsFe8. Accordingly, the N\'eel vector tunnels back
and forth at least eight times before the oscillation decoheres. MQC in CsFe8 is evident.

\begin{figure}
\includegraphics{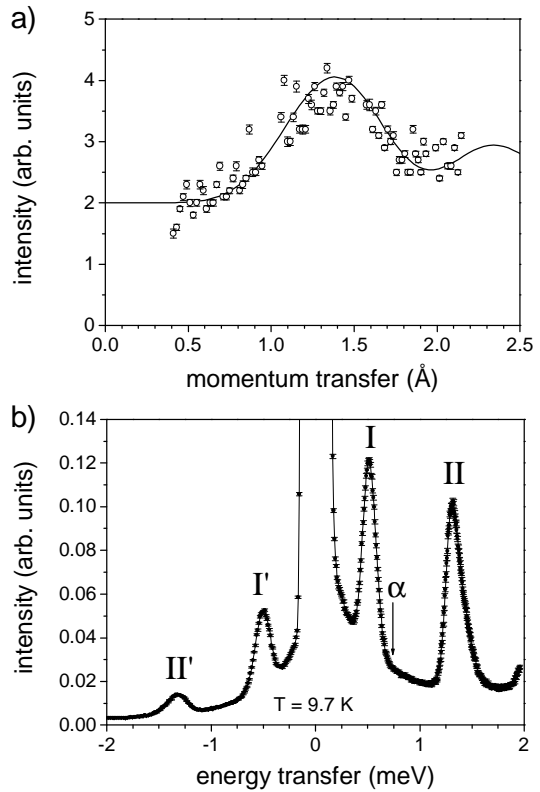}
\caption{\label{fig4} (a) Dependence of the intensity of peak I on momentum transfer $Q$. The solid line
represents the theoretical curve for a $\Delta q = \pi$ transition. The analytical result was scaled, and a
constant background accounting for the incoherent scattering in non-deuterated samples was added. (b) Neutron
scattering intensity of CsFe8 at $T$ = 9.7~K ($\lambda$ = 5.0~\AA).}
\end{figure}

Further details of the INS measurements are noteworthy. The $Q$ dependence of the intensity of peak I was
extracted from the $\lambda$ = 5.0~\AA, $T$ = 2.4~K data by fitting a Gaussian with sloping background to the
data of each detector bank. The result is displayed in Fig.~\ref{fig4}(a). The $Q$ dependence probes directly
the spatial symmetry properties of the involved wave functions \cite{INS}. The cyclic symmetry of the
molecular wheels implies a shift operator $\hat{T}$, which moves spins by one position from $i$ to $i + 1$.
This produces a shift quantum number $q$ via $\hat{T}|q\rangle = e^{iq}|q\rangle$, by which the eigenstates
can be classified ($q = 2\pi n/N$ with $n = 0, \ldots, N - 1$) \cite{Sym}. The special values $q = 0$ and $q
= \pi$ correspond to states which are even and odd under inversion, respectively. For molecular wheels, the
$Q$ dependence can be calculated analytically and is found to depend only on the difference in the shift
quantum numbers, $\Delta q = |q - q'|$. The intensity of a  $\Delta q = \pi$ transition is obtained as
$I_{\Delta q = \pi} \propto f^0_{04}(Q) + f^2_{04}(Q)$, with the interference terms $f^0_{04}$ and $f^2_{04}$
given by eq.~(16) of Ref.~\onlinecite{INS}. The observed $Q$ dependence is in good agreement with the
theoretical expectation, proving that peak I corresponds to a transition between states with even and odd
parity, exactly as is expected for a transition between the two tunneling states $|\uparrow\rangle +
|\downarrow\rangle$ and $|\uparrow\rangle - |\downarrow\rangle$.

INS spectra were also recorded at $T$ = 9.7~K [Fig.~\ref{fig4}(b)]. Peaks I and II, as well as their
counterparts on the neutron energy-gain side, I' and II', are clearly observed. The first excited level has
significant thermal population at this temperature, and a transition from the first to the second excited
level (transition $\alpha$) is expected at 0.80~meV on the loss side. Remarkably, this transition is not
observed in the experiment. This can be related to sublattice magnetizations of mesoscopic length in CsFe8.
In a simplifying argument, the relative intensities of the transitions are essentially governed by the
squares of the Wigner-9$j$ symbols
\begin{equation}
\nonumber
W(S_A,S_B,S,S') \equiv \left\{ \begin{array}{ccc} S_A&S_B&S \\
S_A&S_B&S' \\ 0&1&1  \end{array} \right\},
\end{equation}
where $S_A = S_B = Ns/2$ denote the total spin on each AF sublattice ($S = 0$, $S' = 1$ for transitions I and
II, $S = S' = 1$ for transition $\alpha$) \cite{alpha}. For a microscopic system such as a dimer of $s = 1$
ions where $S_A = S_B = 1$ and $W^2(1,1,1,1) / W^2(1,1,0,1) = 1/4$, the three transitions are of comparable
intensity. For CsFe8, in contrast, with its mesoscopic sublattice magnetization $M_0 = 20\mu_B$,
corresponding to $S_A = S_B = 10$ and $W^2(10,10,1,1) / W^2(10,10,0,1) = 1/220$, transition $\alpha$ is two
orders of magnitude weaker than transitions I and II. The observed weak intensity of transition $\alpha$ in
CsFe8 is thus a direct experimental signature for the tunneling of a mesoscopic variable.

%

Our findings conclusively demonstrate coherent NVT in CsFe8, providing access to entangled quantum
superpositions of the mesoscopic states $|\uparrow\rangle = |M_0,-M_0\rangle$ and $|\downarrow\rangle =
|-M_0,M_0\rangle$. For a Mn4 dimer and a Ni4 molecule, decoherence times of $\tau_\phi \gtrsim 1$~ns were
inferred recently \cite{Hil04,Bar04}. As AF particles with compensated magnetization, such as CsFe8, are
expected to couple less strongly to the environment \cite{Bar90}, the actual decoherence times in such
particles might be much longer. For CsFe8 this implies $\tau_\phi \gg 1$~ns, or $\tau_\phi \Delta/\hbar \gg
10^3$, respectively. The threshold of $\tau_\phi \Delta/\hbar > 10^4$ for potential applications seems thus
reachable, identifying AF molecular wheels as promising prototypes for realizing AF cluster qubits
\cite{Mei03}. Since the original suggestion of MQC in AF molecular wheels \cite{Chi98,Mei01}, a large number
of different wheels have been synthesized and magnetically characterized; yet only CsFe8 exhibits a $k_z/J$
ratio large enough for NVT to occur in zero magnetic field. The synthesis of further molecular wheels
exhibiting NVT remains a challenge, but is highly desirable.

%

\begin{acknowledgments}
We are indebted to F. Meier for numerous helpful discussions on the subject. Financial support by
EC-RTN-QUEMOLNA, contract n$^\circ$ MRTN-CT-2003-504880, and the Swiss National Science Foundation is
acknowledged.
\end{acknowledgments}

%

%

\begin{references}
\bibitem{dip}
Intramolecular dipole-dipole interactions cannot be neglected, but have similar effects as the single-ion
anisotropy and are well grasped by $\hat{S}^2_{iz}$ terms. The $k_z$ value should be thus understood as to
parameterize both anisotropy contributions.
\bibitem{kz}
In principle, $k_z$ could be also negative, but the AF wheels characterized so far exhibit $k_z > 0$. For
CsFe8, $k_z > 0$ is independently confirmed by high-field torque and susceptibility measurements
\cite{CsFe8}, and by an analysis of the intensities of the INS peaks I and II.
\bibitem{crit}
Ref.~\onlinecite{Mei01} gives the criterion $k_z/J \gtrsim 2(Ns)^{-2}$ or $S_0/\hbar \gtrsim 2$ for the
crossover to the NVT regime in zero magnetic fields, but considers that NVT requires that the height of the
anisotropy barrier $E_B = N s^2 k_z$ is larger than the ground state energy $\hbar \omega_0 = s
\sqrt{8Jk_z}$. This provides the stringent, and in our view more conclusive, criterion $S_0/\hbar > 4$.
\bibitem{size}
The tunneling gap could be derived analytically in the case of large magnetic fields perpendicular to the $z$
direction \cite{Chi98,Mei01}. Here, the tunneling gap shows a oscillation as function of field with an
amplitude of $\Delta_{max} = 8 \hbar \omega_0 \sqrt{S_0/(2\pi\hbar)} \exp(-S_0/\hbar)$. Numerical
calculations \cite{FW_QT,Hon02} indicate that the size of the tunneling gap in zero field, $\Delta$, is ca.
twice as large as $\Delta_{max}$, yielding $\Delta \approx 2 \Delta_{max}$.
\bibitem{broad}
The observed inhomogeneous broadening is associated with a distribution in $J$ and $k_z$ values of the
molecules in the sample, most likely due to small structural distortions resulting from an unavoidable loss
of solvent molecules during the preparation of powder samples.
\bibitem{alpha}
The presented argument is strictly valid for small anisotropies, and thus inconsistent with the NVT scenario
found in CsFe8. A more involved argument for very large anisotropies arrives at the same conclusion.
Detailed numerical studies in the full range of anisotropies showed that the relative intensities exhibit
little variation in terms of orders of magnitude. The argument for small anisotropies was chosen here for
its transparency.
\end{references}
\end{document}